\documentclass[12pt]{article}

\def\half{\textstyle{\frac{1}{2}}}
\def\quarter{\textstyle{\frac{1}{4}}}

\def\ra{\rightarrow}
\def\s{\hskip.08em}
\def\tint{{\textstyle\int}}
\def\E{{\rm E}\hskip-.55em{\rm I}\hskip.2em}
\def\hphi{\hat{\phi}}
\def\hpi{\hat{\pi}}
\def\d{\partial}
\def\H{{\cal H}}
\let\frak\bf
\def\l{\lambda}

\def\b{\begin{eqnarray*}}      
\def\e{\end{eqnarray*}}        
\def\bn{\begin{eqnarray}}         
\def\en{\end{eqnarray}}           
\def\<{\langle}
\def\t{\tau}
\def\>{\rangle}

\def\{{\lbrace}
\def\}{\rbrace}
\title{Ultralocal Fields and their Relevance for\\
Reparametrization Invariant\\ Quantum Field Theory}

\author{John R. Klauder\\
Departments of Physics and Mathematics\\
University of Florida\\
Gainesville, FL  32611\\
email: klauder@phys.ufl.edu}
\date{}          
\begin{document}
\maketitle
\begin{abstract}
 Reparametrization invariant theories have a vanishing Hamiltonian and 
enforce their dynamics through a constraint. We specifically choose the 
Dirac procedure of quantization before the introduction of constraints. 
Consequently, for field theories, and prior to the introduction of any 
constraints, it is argued that the original field operator representation 
should be ultralocal in order to remain totally unbiased toward those 
field correlations that will be imposed by the constraints. It is shown 
that relativistic free and interacting theories can be completely recovered 
starting from ultralocal representations followed by  a careful 
enforcement of the appropriate constraints. In so doing all unnecessary 
features of the original ultralocal representation disappear. 

The present discussion is germane to a recent theory of affine quantum 
gravity in which ultralocal field representations have been invoked 
before the imposition of constraints.
\end{abstract}
\vfill\eject
\section{Introduction}
\subsubsection*{Basic classical model}

A variety of theories enjoy invariance under reparametrization of 
the time
variable, and this feature represents one aspect of a favored goal of
achieving invariance under arbitrary coordinate transformations. We have
in mind theories as complex as general relativity or as simple as a single
degree of freedom dynamical system expressed in a reparametrization 
invariant formulation. For a single degree of freedom system, let the 
initial variables of the canonical formulation be called $p$ and $q$. 
We note that the physics of the situation typically dictates the 
dimensions of the variables $p$ and $q$, and occasionally the dimensions 
of the momentum and coordinate variables will  be of interest. 
We generally describe a simple dynamical system by the action functional
  \b   I=\tint[p\,{\dot q}-H(p,q)]\,dt\;, \e
where ${\dot q}=dq/dt$, and the equations of motion it gives rise to, 
viz., 
 \b \frac{dq}{dt}=\frac{\d H}{\d p}\;,\hskip1cm \frac{dp}{dt}=
- -\frac{\d H}{\d q} \;, \e
subject to suitable boundary data. 

By a reparametrization invariant formulation we have in mind the 
alternative action functional
\b I'=\tint\{p\,(dq/d\t)+s\,(dt/d\t)-\l(\t)[s+H(p,q)]\}\,d\t\;. 
\e
In going from $I$ to $I'$ we have introduced a new independent 
variable $\t$, elevated the former independent variable $t$ to a 
dependent (dynamical) variable, and introduced its conjugate variable 
$s$. Additionally, we note that in the new form the Hamiltonian vanishes, 
while in its place we have a Lagrange multiplier variable $\l(\t)$ which 
enforces the (first-class) constraint $s+H(p,q)=0$. Based on these facts, 
the equations of motion that arise from $I'$ read as 
\b \frac{dq}{d\t}=\l\frac{\d H}{\d p}\,,\hskip.5cm \frac{dp}{d\t}=
- -\l\frac{\d H}{\d q}\,,\hskip.5cm\frac{dt}{d\t}=\l\,,\hskip.5cm 
\frac{ds}{d\t}=0 \e
along with the constraint $s+H(p,q)=0$. From these equations we 
deduce that $dq/d\t=(dt/d\t)(\d H/\d p)$, etc., i.e., 
 \b \frac{dq}{dt}=\frac{\d H}{\d p}\;,\hskip1cm \frac{dp}{dt}=
- -\frac{\d H}{\d q}\;,  \e
exactly as before. 

Observe, in the former case, that the dynamics arose from the 
{\it Hamiltonian}; in the latter case, the dynamics arose from 
enforcing a constraint, that is, it arose from the {\it kinematics}.
\subsubsection*{Basic quantum model}
A quantum analysis of reparametrization invariance follows the spirit 
of the preceding discussion. The initial quantum theory assumes 
canonical, self-adjoint operators $P$ and $Q$, with $[Q,P]=i$ (in 
units where $\hbar=1$), and a Hamiltonian operator $\H=\H(P,Q)$ all 
acting in a Hilbert space $\frak H$. The operator equations of motion 
are given by
 \b \frac{dQ(t)}{dt}=i[\H,Q(t)]\;,\hskip1cm\frac{dP(t)}{dt}=i[\H,P(t)]\;, 
\e
and expectations of interest are given, say, by
 \b  \<\psi|\,Q(t_1)\,Q(t_2)\,\cdots\,Q(t_n)\,|\psi\>\;, \e
etc., for a general $|\psi\>\in{\frak H}$. 

As another characterization of the theory, we may also employ a coherent 
state representation \cite{klsk}. To that end let us introduce
  \b  |p,q\>\equiv e^{-iqP}\,e^{ipQ}\,|\eta\>   \e
where $|\eta\>$ denotes a suitable normalized fiducial vector which we 
generally choose as an oscillator ground state defined by 
$(\Omega Q+iP)\,|\eta\>=0$. 
In that case,
  \b &&\hskip-.5cm\<p'',q''|p',q'\>\\
&&\hskip.5cm= \exp{\{\half i(p''+p')(q''-q')-
\quarter[\Omega^{-1}(p''-p')^2+\Omega(q''-q')^2]\}}\;.  \e
Here, $\Omega>0$ is a free parameter of the coherent state 
representation and  may be chosen arbitrarily. Under normal 
circumstances, a parameter such as $\Omega$ is necessary on 
dimensional grounds, but even if it is
dimensionless it is still a freedom in the coherent state 
representation. For simplicity, it is often convenient to set 
$\Omega=1$; however, we shall not do so. It is important to observe 
that the coherent state overlap 
$\<p'',q''|p',q'\>$ defines a positive-definite function which we may 
choose as a {\it reproducing kernel} and use it to define the associated 
{\it reproducing kernel Hilbert space} \cite{mesh}. With regard to 
dynamics for this system,
we may study the propagator implicitly given by
  \b  \<p'',q''|\,e^{-i(t''-t')\H}\,|p',q'\>\;.  \e
Hereafter, we shall principally focus on coherent state formulations.

For the reparametrization invariant formulation, on the other hand, the 
quantization procedure is somewhat different. Adopting Dirac's method 
\cite{dir}, we proceed as follows. Let $P,Q$ and $S,T$ denote two 
independent canonical pairs with $[S,T]=i$ as the only additional 
nonvanishing commutator. Thus the Hilbert space ${\frak H}'$ is now 
that for a two degree-of-freedom problem. To reduce that space to one 
appropriate to a single degree of freedom, we (ideally) select out a 
subspace ${\frak H}_{phys}\subset{\frak H}'$ by the constraint condition
 \b  [S+\H(P,Q)]\,|\psi\>_{phys}=0\;.  \e

It is especially convenient to  choose a coherent state approach in 
which to enforce this kind of constraint \cite{kla1}. To that end we 
introduce the coherent state
  \b  |p,q,s,t\>= e^{-iqP}\,e^{ipQ}\,e^{-itS}\,e^{isT}\,|\eta'\>=
|p,q\>\otimes|s,t\>  \e
in the Hilbert space ${\frak H}'$.
The overlap of two such coherent states is given (with $F$ an evident 
normalization factor) by
  \b &&\hskip-1cm \<p'',q'',s'',t''|p',q',s',t'\>=\<p'',q''|p',q'\>
\<s'',t''|s',t'\>  \\
      && =F\int\int e^{-(k-p'')^2/2\Omega-(u-s'')^2/2\Lambda} \\
       &&\hskip1.4cm  \times e^{i(q''-q')k+i(t''-t')u} \\
       &&\hskip1.4cm\times e^{-(k-p')^2/2\Omega-(u-s')^2/2\Lambda}
\,dk\,du  \e
We impose the constraint in this expression in the form
 \b &&\<p'',q'',s'',t''|\,\delta(S+\H(P,Q))\,|p',q',s',t'\> \\
&&\hskip.5cm\equiv\lim_{\delta\ra0}\,(2\delta)^{-1}\,
\<p'',q'',s'',t''|\,\E([S+\H(P,Q)]^2\le\delta^2)\,|p',q',s',t'\>\;,  \e
where $\E$ is a projection operator onto the spectral interval 
$-\delta\le [S+\H(P,Q)]\le\delta$, which, 
up to a multiplicative factor, leads to 
 \b &&\hskip-1cm \<\!\<p'',q'',s'',t''|p',q',s',t'\>\!\> \\
     && \equiv \<p'',q'',0,0|\,e^{-(s''+\H(P,Q))^2/2\Lambda}\,
e^{-i(t''-t')\H}\,e^{-(s'+\H(P,Q))^2/2\Lambda}\,|p',q',0,0\>  \;.  \e  
At this point the variables $s$ and $t$ do not add to the span of the 
reduced Hilbert space. Or, to put it another way, the sets of states 
$\{|p,q,s,t\>\!\>\}$ and $\{|p,q,0,0\>\!\>\}$ span the same Hilbert 
space. Therefore, we can for example integrate out the variables 
$s''$ and $s'$ without changing matters, which, with a suitable 
rescaling, then leads to the positive definite function
 \b \<\!\<p'',q'',t''|p',q',t'\>\!\> 
      \equiv \<p'',q'',0,0|\,e^{-i(t''-t')\H}\,|p',q',0,0\>\;. \e
Finally, we may simply identify the last result with the same expression 
we obtained previously,
that is,
  \b \<\!\<p'',q'',t''|p',q',t'\>\!\>=\<p'',q''|\,
e^{-i(t''-t')\H}\,|p',q'\>\;,
\e
completing the argument that the reparametrization formulation leads 
to the same result as the original formulation. This example also 
illustrates how ``time'' may emerge within a reparametrization 
invariant formulation.

\subsubsection*{Discussion}
The foregoing examples have been limited to a single degree of freedom. 
It is clear, however, that an analogous discussion holds in the case of 
finitely many degrees of freedom. While the parameter $\Lambda$ has 
disappeared from the final fully constrained theory, the parameter 
$\Omega$ of the coherent state representation remains. However, it 
merely corresponds to an inherent freedom in the representation that 
has no physical consequence whatsoever.

The situation changes, however, when an infinite number of degrees of 
freedom are present. As is well known, and unlike the situation for 
finitely many degrees of freedom, fields that satisfy canonical 
commutation relations possess uncountably many inequivalent irreducible 
representations, and it is imperative that one choose the correct one in 
order for a field theory to make sense \cite{haag}. Prior to the 
introduction of the constraint(s), there is in principle no information 
{}from which to make a proper choice of field operator representation. In 
fact, before any constraints have been introduced, one must essentially 
turn a blind eye toward what constraints are to come and initially use a 
representation that is as neutral toward the situation as is possible. 
With that thought in mind, we are led to propose that the only neutral 
field operator representations that should be considered are 
{\it ultralocal representations}. Ultralocal representations are defined 
as those that have {\it no} correlations between fields at distinct 
spatial points, i.e., fields whose values at distinct spatial points 
are statistically independent. Any correlations between fields at 
distinct spatial points that are required by the physics will be 
introduced by the constraints that are imposed later. (Alternative 
quantization procedures which reduce first and quantize second can 
lead to erroneous results \cite{SHAB}.)

Having argued that, before the introduction of constraints, the initial 
field operator representation should be ultralocal, leads to a possible 
dilemma since the ultralocal field operator representations are surely 
inequivalent to the desired field operator representations that are 
expected once the constraints are fully enforced. The question arises 
as to whether this inequivalence, or more descriptively, this huge 
distance between representations can be bridged, and, even if it can, 
will there be any unwelcome residue remaining of the original and 
unphysical ultralocal representation. Happily, as we shall learn, it 
is entirely possible to start with an ultralocal representation, as an 
unbiased starting point requires, and, nevertheless, emerge with the 
proper representation for the problem at hand---and moreover, no 
unwelcome residue of the original representation remains. In the sense 
described, ultralocal field theories appear to have found genuine 
physical applications in the quantization of reparametrization invariant 
field theories. 

Indeed, elsewhere \cite{kla2}, we
have already discussed the important role ultralocal fields play in the 
quantization of the gravitational field by affine field variables before 
any constraints are introduced; the present paper may be considered as 
an adjunct to this same program for quantum gravity.

In principle, there is one marked advantage in starting with a 
reparametri- zation invariant formulation and the associated ultralocal 
representations of the fields that are involved. The advantage of doing 
so arises from the fact that {\it all ultralocal representations can be 
classified, and they already have been rather thoroughly investigated} 
\cite{klabook} (see Sec.~3 below). 
 
\subsubsection*{Remaining sections}
In the following section, Sec.~2, we shall extend the analysis of the 
present section to several different field theories: relativistic free 
fields, and relativistic $\phi^4_2$ and $\phi^4_3$ models. In Sec.~3, 
we shall review the most general ultralocal field operator 
representations, briefly revisit our discussion in Sec.~2, and just 
mention the case of ultralocal model fields themselves and how they 
fit into the general scheme.

Speaking more generally, we will argue in the following sections that 
even though ultralocal fields are initially used, the usual (and 
generally nonultralocal) results still emerge from the reparametrization 
invariant formulation after the constraints have been fully enforced. 

\section{Relativistic Fields}
\subsection{Conventional free fields}
\subsubsection*{Classical theory}
A classical relativistic free field $\phi=\phi(x,t)$ [and momentum 
$\pi=\pi(x,t)$] of mass $m$ in (say) four space--time dimensions may 
be described by the action functional
  \b I=\tint dt\tint d^3\!x\{\pi\,{\dot\phi}-\half[\pi^2+
(\nabla\phi)^2+m^2\phi^2]\}\;, \e
which leads to the equations of motion $\d\phi/\d t\equiv{\dot\phi}=
\pi$ and
 \b {\dot\pi}={\ddot\phi}=\nabla^2\phi-m^2\phi \;. \e
Given suitable initial conditions, solution of these equations of 
motion follows conventional lines. 
\subsubsection*{Quantum theory}
Canonical quantization proceeds by introducing irreducible, locally 
self adjoint field ${\hat\phi}(x)$ and momentum ${\hat\pi}(x)$ operators 
that satisfy the canonical commutation relations 
  \b [{\hat\phi}(x),{\hat\pi}(y)]=i\delta(x-y)\;,  \e
with all other commutators vanishing. Of all the (inequivalent) operator 
representations that fulfill the commutation relations, attention is 
focussed on that one which admits
 \b \H\equiv\half\tint:\{{\hat\pi}(x)^2+[\nabla{\hat\phi}(x)]^2+
m^2{\hat\phi}(x)^2\}:\,d^3\!x  \e
as a nonnegative, self-adjoint operator, and for which $|0\>$ denotes 
a nondegenerate, normalized ground state with $\H\,|0\>=0$. Here the 
colons $:\;\;:$ signify normal ordering with respect to $|0\>$. The 
chosen representation of ${\hat\phi}(x)$ and ${\hat\pi}(x)$ depends 
on $m$ and is unitarily inequivalent to any other representation 
defined for a value $m'\ne m$. 

We adopt a coherent state representation for this example (see, e.g., 
\cite{klsu}), and we therefore introduce
 \b  |\pi,\phi\>\equiv e^{i[{\hat\phi}(\pi)-{\hat\pi}(\phi)]}\,|0\> \e
where ${\hat\phi}(\pi)=\tint {\hat\phi}(x)\s\pi(x)\,d^3\!x$ and 
$\hpi(\phi)=\tint\hpi(x)\s\phi(x)\,d^3\!x$ are
defined for smooth (test) functions $\pi$ and $\phi$. (We have 
adopted a different phase convention for the coherent states as 
compared to that in Sec.~1, but this is of no significance.) In turn, 
the propagator in this representation is expressed by
  \b \<\pi'',\phi''|\,e^{-i(t''-t')\H}\,|\pi',\phi'\>  \e
and it has the explicit expression given by
  \b \<\pi'',\phi''|\,e^{-i(t''-t')\H}\,|\pi',\phi'\> =N''N'
\exp[\tint z''^{*}(k)\,e^{-i(t''-t')\omega(k)}\,z'(k)\,d^3\!k]\;,  \e
where in each case $z(k)\equiv(1/\sqrt{2})[\omega(k)^{1/2}{\tilde 
\phi}(k)+i\s\omega(k)^{-1/2}{\tilde \pi}(k)]$, and ${\tilde \pi}(k)$ 
and ${\tilde \phi}(k)$ are the Fourier transform of $\pi(x)$ and 
$\phi(x)$, respectively. The factors $N''$ and $N'$ ensure 
normalization and are given, in each case, by
  \b  N=\exp[-\half\tint|z(k)|^2\,d^3\!k]\;,  \e
and finally $\omega(k)\equiv(k^2+m^2)^{1/2}$.  

It is noteworthy that the mass parameter $m$ labels {\it mutually 
distinct} 
Hilbert spaces with the property that any vector $|\psi;m\>\in{\frak H}_m$
and any vector $|\chi;m'\>\in{\frak H}_{m'}$, $m\ne m'$, are 
{\it orthogonal}. Another way of saying this is that the field operator 
representations of $\hat\phi$ and $\hat\pi$ are unitarily inequivalent 
for any two mass values $m$ and $m'$ whenever $m\ne m'$. We stress these 
facts to emphasize that quantum theories of relativistic free fields are, 
for different mass values, fundamentally distinct from one another; if you 
have one of them, you definitely do {\it not} have any of the others!

\subsection{Reparametrization invariant free fields}
\subsubsection*{Classical theory}
In analogy with the simple example of Sec.~1, we introduce the 
reparametrization invariant action for the relativistic free field 
of mass $m$ given by
 \b I'=\tint d\tau(\!\!(st^*+\tint d^3\!x\,[\pi\phi^*]-\lambda\{s+
\half\tint d^3\!x\,[\pi^2+(\nabla\phi)^2+m^2\phi^2]\})\!\!)\;,  \e
where $t^*\equiv d t/d\tau$, $\phi^*\equiv \d\phi/\d\tau$, and $\lambda=
\lambda(\tau)$ denotes a Lagrange multiplier function. It is clear that 
the equations of motion that follow from $I'$ lead to the same equations 
of motion that follow from $I$, just as was the case for the simple 
example in Sec.~1.
 
\subsubsection*{Quantum theory}
Once again we introduce coherent states for the combined system given by
\b |\pi,\phi,s,t\>\equiv |\pi,\phi\>\otimes|s,t\>\;,  \e
where in the present case we adopt an ultralocal representation for the 
field
operators since we assume prior ignorance of the constraints that are 
yet
to be introduced. As a consequence, the overlap of two such coherent states 
is given by
  \b  &&\<\pi'',\phi'',s'',t''|\pi',\phi',s',t'\>  \\
  &&\hskip1cm =L''L'\exp[\tint u''^{*}(x)\,\,u'(x)\,d^3\!x] \\
   &&\hskip1.3cm\times\exp\{-\quarter[\Lambda^{-1}(s''-s')^2-2i(s''+s')
(t''-t')+\Lambda(t''-t')^2]\}\;,  \e
where in the present case
  \b  &&u(x)\equiv (1/\sqrt{2})[M^{1/2}\phi(x)+i\s M^{-1/2}\pi(x)]\;, \\
      &&\hskip.53cm L\equiv \exp[-\half\tint|u(x)|^2\,d^3\!x]  \e
in both of the cases $u=u''$ and $u=u'$. Here $M$, $0<M<\infty$, is a 
new parameter in the ultralocal representation
which labels {\it inequivalent} field operator representations for 
$\hat{\pi}$ and $\hat{\phi}$, representations which in addition are 
inequivalent to all relativistic free field representations for any mass 
$m$. For the usual free fields that we are considering in this section, 
$M$ must have the dimensions of ``mass'', and thus $M$ is necessary simply 
on dimensional grounds if for no
other reason. Like the parameter $\Omega$ in the simple example of Sec.~1,
we are free to choose it; but in the present case, since it labels 
inequivalent field operator representations, it would appear that the 
choice of $M$ is by no means benign. 

In contrast to such appearances, we shall now show that one may start 
with {\it any} choice of $M>0$ and emerge with the usual quantization 
of the relativistic free field for {\it any} (independent) choice of 
$m\ge0$ (with, as usual, extra care needed when $m=0$ if the space were 
one dimensional instead of three). In fact, when the constraint is fully 
enforced, we will find that {\it all traces of the arbitrary parameter 
$M$ will disappear from the final result}.

As in the simple example of Sec.~1, we now proceed to introduce the 
constraint. However, the operator form of the quantum constraint 
strongly depends on having the ``right'' field operator representation. 
Stated otherwise, if we build 
  \b  \H=\half\tint:\{\hpi(x)^2+[\nabla\hphi(x)]^2+m^2\hphi(x)^2\}:
\,d^3\!x  \e
out of the present field operator representation plus normal ordering 
with respect to the fiducial vector (or any vector for that matter), we 
will find that it is {\it not an operator}, or more precisely that it 
has only the zero vector in its domain. To proceed, let us first introduce 
a regularized form of the operators appearing in the constraint. To that 
end, let $\{h_n(x)\}_{n=1}^\infty$ denote a complete orthonormal set of 
real functions over the configuration space. For each 
$N\in\{1,2,3,\ldots\}$, define the sequence of kernels
  \b  K_N(x,y)\equiv\sum_{n=1}^N\,h_n(x)\,h_n(y)  \;,  \e
which have the property that, as $N\ra\infty$, $K_N(x,y)\ra\delta(x-y)$, 
in the sense of distributions. Next consider the regularized fields
 \b && \hpi_N(x)\equiv \tint K_N(x,y)\,\hpi(y)\,d^3\!y\;,  \\
    && \hphi_N(x)\equiv \tint K_N(x,y)\,\hphi(y)\,d^3\!y\;.  \e
Now build the regularized Hamiltonian operator
\b\H_N\equiv\half\tint:\{\hpi_N(x)^2+[\nabla\hphi_N(x)]^2+m^2\hphi_N(x)^2\}:
\,d^3\!x\;,\e
which also involves normal ordering with respect to the fiducial vector.
Observe, for $N<\infty$, that the operator $\H_N$ involves only a finite 
number of degrees of freedom. 

We are now in position to enforce the regularized constraint by considering
 \b \<\pi'',\phi'',s'',t''|\,\delta(S+\H_N)\,|\pi',\phi',s',t'\>\;, \e
which, as in Sec.~1, up to a finite coefficient, assumes the form
  \b \<\pi'',\phi'',0,0|\,e^{-(s''+\H_N)^2/2\Lambda}\,e^{-i(t''-t')\H_N}
\,e^{-(s'+\H_N)^2/2\Lambda}\,|\pi',\phi',0,0\> \;.  \e
In this form we again observe that the parameters $s$ and $t$ do not lead 
to any new vectors in the reduced Hilbert space. In other words, $s$ and 
$t$ no longer play a role in spanning the space of vectors. While
this expression is well defined for all $N<\infty$, it will vanish 
identically in the limit $N\ra\infty$ for any choice of $m$ (because, 
in effect, $\H_N\ra\infty$). That is one signal that the present 
(ultralocal) representation is {\it not} well suited to the relativistic 
free field. This fact is of course not surprising, but it is comforting 
to have an unambiguous analytic signal of this incompatibility. 

It does not help to integrate out the variables $s''$ and $s'$.  The 
result of that integration, apart from finite factors, is given by
\b \<\pi'',\phi'',0,0|\,e^{-i(t''-t')\H_N}\,|\pi',\phi',0,0\>\;, \e
which is well defined and continuous in its arguments when $N<\infty$, but
is {\it no longer} continuous in the time variable after taking the limit 
$N\ra\infty$. Specifically, the limit is well defined when $t''=t'$, but 
it is undefined when $t''\ne t'$. This is yet another concrete signal that 
the ultralocal representation is not well suited to the relativistic free 
field. 

As already stated, these unsuitable limits have emerged because we do not 
have the right field operator representation. However, we can fix that. Let 
us first make a selection of which relativistic free field we are 
interested in, i.e., let us choose a mass $m$. Then, by taking suitable 
linear combinations of the previous expression we can build a new 
expression given by 
 \b \<\pi'',\phi'';N,m|\,e^{-i(t''-t')\H_N}\,|\pi',\phi';N,m\>\;, \e
which involves coherent states that are {\it not} based on the original 
fiducial vector $|\eta\>$, but, instead, are based on the fiducial 
vector $|N,m\>$ which for the first $N$ degrees of freedom is defined 
to be the ground state of the operator $\H_N$, and is unchanged for the 
remaining degrees of freedom. We suppose further that $\H_N$ is also 
normal ordered with respect to the vector $|N,m\>$. It is now clear that 
we can indeed take the limit $N\ra\infty$ and obtain an expression which 
is continuous in the time variable in the limit. In this manner we have 
effectively passed from the wrong (ultralocal) representation of the 
field operators to the right (relativistic) representation of the field 
operators. After $N\ra\infty$, the result becomes
 \b \<\pi'',\phi''|\,e^{-i(t''-t')\H}\,|\pi',\phi'\>\;,  \e
where 
 \b \H=\half\tint:\{\hpi(x)^2+[\nabla\hphi(x)]^2+m^2\hphi(x)^2\}:\,d^3\!x  
\e
is now well defined and coincides with the usual free field Hamiltonian. 
Therefore, the final propagator {\it completely coincides} with the 
expression we previously found with the same name. This argument 
demonstrates that the reparametrization invariant formulation leads 
to the same result as the usual theory even though we started from an 
ultralocal formulation of the unconstrained theory. Observe, in addition, 
that all traces of the initial parameter $M$ have disappeared in the final 
expression. In short, whatever $M$ value had been originally chosen, it 
served only as a ``place holder'' for the proper expression (effectively, 
$\omega(k)$, which it should be noted has the same dimensions as $M$) 
that emerged when the constraint was 
fully enforced.

There may well be some questions remaining about how one may attain 
the ``suitable linear combinations'' to secure our desired result. To 
that end we offer the following rather picturesque description of how 
that may be accomplished. Consider the set of inner product quotients 
of a dense set of 
vector norms given by
\b  S\equiv\bigg\{\;\frac{\Sigma_{j,k=1}^J\,a^*_j\,a_k\,
\<\pi_j,\phi_j,0,0|\,e^{-\H_N^2/\Lambda}\,|{{\pi}}_k,{{\phi}}_k,0,0\>\;}
{\Sigma_{j,k=1}^J\,a^*_j\,a_k\,
\<\pi_j,\phi_j,0,0|{{\pi}}_k,{{\phi}}_k,0,0\>\;}:\;J<\infty\;\bigg\} \;,\e
where $\{a_j\}$ denote complex coefficients (not all zero), while 
$\{\pi_j\}$ and $\{\phi_j\}$ denote suitable fields. (The numerator of 
this quotient arises from the original expression when $s''=s'=0$ and 
$t''=t'$; the denominator of this quotient arises after integrating out 
the variables $s''$ and $s'$ when $t''=t'$.) Now, as $N$ becomes large, 
it will happen that most of the inner product quotients in the set $S$ 
will become exponentially small, while some will not. The vectors whose 
quotients do not become small lie in the general direction in the 
pre-Hilbert space where the proper ground state has the largest overlap 
with the present set of vectors. One then takes the necessary linear 
combination of vectors, and any necessary limits, to change the fiducial 
vector to the unit vector that maximizes this overlap. Introducing a new 
set of coherent states that has the maximizing vector as its fiducial 
vector is a process that may be called {\it recentering the coherent states} 
or equivalently {\it recentering the reproducing kernel}. It is this 
procedure that ``homes in'' on the right representation of the field 
operators for the relativistic free field, and ultimately leads to a 
suitable and continuous expression for the reproducing kernel for the 
relativistic free field. It is clear in this construction that in the 
limit all traces of the original ultralocal representation disappear and 
one is left only with the proper relativistic expression. 

In summary, we note that regularization of the constraint and careful 
recentering of the coherent states as the regularization is removed are 
the key procedures in passing from the form of the expression before the 
constraint has been introduced to the form in which the constraint has 
been fully enforced. Equivalence between the usual and reparametrization 
invariant formulations has thus been established, which was our goal.

\subsection{Other relativistic fields}
The preceding discussion was confined to relativistic free fields, but we 
may also consider interacting fields as well. In particular, the 
so-called $\phi^4_2$ and $\phi^4_3$ models have been rigorously 
constructed \cite{bau}. Both of these models admit canonical field 
and momentum operators that obey canonical commutation relations. In 
addition, both models admit unique ground states to their respective 
Hamiltonian operators, but in the present case the ground states are not 
Gaussian in a field diagonal representation as befit truly interacting 
theories. Thus the associated coherent state overlap function of 
interest in the present case is {\it not} quasi-free. Nevertheless, 
we can proceed exactly as before, namely, introduce cutoff fields and 
thereby a cutoff Hamiltonian (with a cutoff dependent bare mass for 
$\phi^4_3$), use that cutoff Hamiltonian to identify a vector 
approximating the ground state of the cutoff Hamiltonian, and then 
recenter the coherent states about a new vector which is based on the 
ground state of the regularized Hamiltonian. Seeking the vector that 
maximizes the elements of the set $S$ will again work in the present 
case as a means to identify the approximate ground state. Taking the 
appropriate limit as the regularization is removed will, just as in the 
free case, lead to a new reproducing kernel and thereby an associated 
reproducing kernel Hilbert space appropriate to the case at hand. Once 
again all traces of the ultralocal representation will disappear in the 
limit that the regularization is removed. Hence, we see that the 
procedures discussed above for the free field case will extend to 
handle those irreducible interacting fields that still satisfy a set 
of canonical commutation relations.

\section{Most General Ultralocal Fields}
In this section we want to discuss more general classes of ultralocal 
field operator representations than those already dealt with in the 
previous section.
We also want to see how these more general representations fit into the 
kind of analysis that was carried out for the relativistic fields in Sec.~1. 

As a first example, we cite the expression for a single field, such as 
the field operator, independent of whether or not it satisfies the 
canonical commutation relations (in the usual sense) with another field. 
In particular, we note that
the most general expression that respects ultralocality is given by
\b  &&\<\eta|\,e^{i\hphi(\pi)}\,|\eta\> \\
   &&\hskip1cm=\exp(\!\!(\tint d^3\!x\,\{\,ia(x)\s\pi(x)-\quarter c(x)
\s\pi(x)^2 \\
&&\hskip2.4cm+\tint[e^{i\lambda\s\pi(x)}-1-i\lambda\s\pi(x)/(1+\lambda^2)]\,
d\sigma(\lambda;x)\})\!\!)\;, \e
where $a(x)$ is an arbitrary real function, $c(x)$ is an arbitrary 
nonnegative function, and $\sigma$ is a nonnegative measure with the 
property that
 \b  \tint[\lambda^2/(1+\lambda^2)]\,d\sigma(\lambda;x)<\infty  \e
for almost all $x$. We do not prove this expression here but rather 
refer the reader to \cite{klabook} where a homogeneous form of this 
result is proved. 
It is straightforward to amend that argument to cover the inhomogeneous case.

In case $c(x)=0$ for a set of nonzero measure, then it is generally 
necessary that $\tint d\sigma(\lambda;x)=\infty$ in that spatial region 
in order that the smeared field $\hphi(\pi)$ has a purely continuous 
spectrum; however, we will not pursue technical issues of this sort here.
\subsubsection*{Canonical fields}
As another class of examples, suppose that the field and momentum 
operators are both involved and that they satisfy conventional canonical 
commutation relations. Then the most general ultralocal representation 
is determined by
\b  &&\<\eta|\,e^{i[\hphi(\pi)-\hpi(\phi)]}\,|\eta\> \\
  &&\hskip1cm=\exp(\!\!(\tint d^3\!x\,\{\,ia(x)\s\pi(x)-ib(x)\s\phi(x)-
\quarter[\, c(x)\s\pi(x)^2 +d(x)\s\phi(x)^2]\\
&&\hskip2.4cm+\tint[e^{i\lambda\s\pi(x)}-1-i\lambda\s\pi(x)/(1+\lambda^2)]\,
d\sigma(\lambda;x) \\    
&&\hskip2.4cm+\tint[e^{-i\gamma\s\phi(x)}-1+i\gamma\s\phi(x)/(1+\gamma^2)]\,
d\rho(\gamma;x)\})\!\!) \;,  \e
subject to the conditions that $a(x)$ and $b(x)$ are arbitrary real 
functions, $c(x)$ and $d(x)$ are positive definite and moreover 
$c(x)\s d(x)\ge1$ holds for almost all $x$, and, finally, $\rho$ 
is a nonnegative measure that fulfills a similar condition as $\sigma$, 
namely 
 \b  \tint[\gamma^2/(1+\gamma^2)]\,d\rho(\gamma;x)<\infty \e
for almost all $x$.

We remark that if either $\sigma$ or $\rho$ is nonzero, or 
if $c(x)\s d(x)>1$ for a nonzero spatial measure, then the field and 
momentum operator representation is reducible rather than irreducible. 
If the expressions $a,\;b,\;c,\;d,\;\sigma$, and $\rho$ are independent 
of $x$, then the representations are homogeneous. Homogenous expressions 
generally suffice to deal with translation invariant constraints, such 
as the examples dealt with in Sec.~2. 

\subsection{Relativistic free field, revisited}
Let us consider the free field once again, and suppose we had chosen a 
different ultralocal representation to start with. In particular, suppose 
we chose
  \b &&\hskip-1.3cm\<\pi'',\phi''|\pi',\phi'\>_a 
    =\exp[\!\![\,\tint d^3x(\!\!(\half i[\phi''(x)\s\pi'(x)-\pi''(x)
\s\phi'(x)] \\
  &&\hskip2.6cm-\quarter\{M^{-1}[\pi''(x)-\pi'(x)]^2+M[\phi''(x)
- -\phi'(x)]^2\}\\
&&\hskip2.6cm+ia(x)[\phi''(x)-\phi'(x)])\!\!)\,]\!\!]\;,  \e
which corresponds to an irreducible representation whatever generalized 
function is chosen for $a$. Recentering the coherent states proceeds as 
in Sec.~2, for any $a$, and results in the proper relativistic coherent 
state overlap for any pregiven mass $m$. Summarizing, we observe that 
recentering the reproducing kernel brings us to the relativistic free 
field form for any choice of $a(x)$. Since that is the case, it follows 
that any suitable superposition of similar expressions over the 
generalized function $a$ will result in the proper free field case. 
In particular, a local, normalized Gaussian superposition over $a(x)$, 
such as
 \b  \int e^{i\tint a(x)[\phi''(x)-\phi'(x)]\,d^3\!x}\,d\mu(a)= 
e^{-\quarter {\tilde M}\tint[\phi''(x)-\phi'(x)]^2\,d^3\!x}\;,  \e
leads to
  \b &&\hskip-.3cm\<\pi'',\phi''|\pi',\phi'\>\equiv \tint
 \<\pi'',\phi''|\pi',\phi'\>_a\,d\mu(a) \\
 &&\hskip2.18cm=\exp[\!\![\,\tint d^3x(\!\!(\half i[\phi''(x)
 \s\pi'(x)-\pi''(x)\s\phi'(x)] \\
  &&\hskip3.4cm-\quarter\{M^{-1}[\pi''(x)-\pi'(x)]^2+
 M'[\phi''(x)-\phi'(x)]^2\})\!\!)\,]\!\!]\;.  \e
In this expression, $M^{-1}M'=M^{-1}(M+{\tilde M})>1$, implying that 
the indicated ultralocal coherent state overlap corresponds to a 
reducible representation of the canonical commutation relations.

The result of this exercise shows that since the recentered reproducing 
kernel 
(after enforcing the constraint)
is given, for any original $a(x)$, by the unique form for the 
relativistic free field, it follows that the superposition of such 
expressions also leads to the relativistic free field form. Stated 
otherwise, we observe that it is even possible to choose an initial 
ultralocal field operator representation that is {\it reducible}, and, 
nevertheless, after regularization and recentering of the coherent 
states, one emerges with the proper {\it irreducible} representation 
for the relativistic free field of mass $m$ in which all traces of the 
original parameters $M$ and $M'\;(>M)$ disappear.

\subsection{Ultralocal model fields}
The last example we consider refers to the so-called ultralocal field 
theories themselves \cite{klabook}. These models only possess sharp time 
local field operators (and not sharp time local momentum operators), and 
thus we recall that
 \b &&\<\pi''|\pi'\>=\<\eta|\,e^{-i[\hphi(\pi'')-\hphi(\pi')]}\,|\eta\> \\
  &&\hskip1.35cm =\exp(\!\!(-b\tint d^3\!x\tint[1-\cos\{\lambda[\pi''(x)
- -\pi'(x)]\}\,c(\lambda)^2\,d\lambda)\!\!) \;, \e
{}$c(-\lambda)^2=c(\lambda)^2$, which is clearly of the general form 
indicated earlier in this section. We focus on the dimensional 
parameter $b$ which appears here as a coefficient. It is noteworthy, 
for these models, that the precise value of $b$ is {\it not} determined 
by the form of the Hamiltonian. In the spirit of the present paper, the 
parameter $b$ is a feature of the original ultralocal representation that 
does {\it not} disappear after the constraint regarding the dynamics is 
fully enforced. We draw from this example the conclusion that while it 
is possible that all traces of the original ultralocal representation may 
disappear once the constraints have been fully enforced, it is not always 
required to be the case. This precaution may be useful to keep in mind in 
studying further examples of reparametrization invariant field theories 
and their quantization using ultralocal representations.

\section*{Acknowledgments}
Thanks are expressed to B. Bodmann for helpful comments.
The work reported in this paper has been partially supported by NSF 
Grant 1614503-12.

\end{document}